\documentclass[12pt]{article}
\usepackage{multicol}
\newcommand{\Vec}[1]{\mbox{\boldmath$#1$}}
\begin{document}
\begin{center}
{\large\textbf{Interaction and dimensionality in the quantum Hall physics}\\ 
Hideo Aoki\\
{\it Department of Physics, University of Tokyo, Hongo,
Tokyo 113-0033, Japan}
}
\end{center}

Abstract. 
While the composite fermion picture is so effective as 
to describe the excitation spectra including 
the spin wave for Laughlin's quantum liquid, 
``how heavy and how strongly-interacting" remains a formidable question 
for the composite fermions, to which this article first addresses.  
The effective mass (purely interaction originated) 
defined from the excitation spectrum and 
obtained for various even- as well as odd-fractions 
exhibits a curious, step-like filling dependence basically 
determined by the number of flux quanta attached to each fermion, 
where the non-monotonic behaviour indicates 
a strong effect of gauge-field fluctuations.  
The excitation spectrum fits a Fermi liquid, but again a large effect of 
inter-composite fermion interaction appears as anomalous Landau's parameters. 

We have then moved on to see how 
the introduction of three-dimensionality (where 
the shape of the Fermi surface becomes relevant) affects 
the interacting electron system, and propose the magnetic-field 
induced SDW in three-dimensional systems. This should be a 
good candidate, in entirely realistic magnetic fields, 
for the integer QHE recently 
predicted by Koshino {\it et al} to 
occur in 3D on the fractal energy spectrum 
similar to Hofstadter's.  
The mechanism for the field-induced 
phase is an effect of interaction in Landau's quantisation 
on incompletely-nested (i.e., multiply-connected) Fermi surfaces, so 
the interplay of many-body physics and the magnetic quantisation 
on various Fermi surfaces may provide an interesting future 
avenue for 3D systems.  

\begin{multicols}{2}
\section{Introduction}
To commemorate the two decades of the quantum Hall effect (QHE), 
I shall discuss two of the important problems 
in the QHE physics --- interaction and dimensionality.  

Fractional quantum Hall system\cite{Pinczuk,Heinonen} 
is unique as a correlated electron 
system in two ways.  First, the system is in the limit of strong 
electron correlation ($U/t$ in the high-$T_C$ language is infinite) 
in that the kinetic energy is quenched 
due to Landau's quantisation.  Second, spatial dimensionality of two
allows Chern-Simons gauge field theoretic treatments.  
The composite fermion picture, one of the most 
fascinating concepts derived from the fractional QHE, 
indeed provides a good description of not only the ground-state 
properties, but even the excitation spectra including 
the spin wave (Fig.1; \cite{nakajima,TMU}) 
and the charge mode\cite{jaincharge} 
for Laughlin's quantum liquid.  

However, we are still far from a complete understanding of the 
composite fermion.   Specifically, as soon as we go beyond the mean field, 
the problem of ``(i) how heavy and (ii) how strongly 
interacting composite fermions are" becomes a formidable question.  
Unlike the ordinary system, Landau's quantisation 
makes the ``dressing the bare mass" impossible.  By the same token 
whether the composite fermions are nearly free or strongly interacting 
is entirely determined by the fluctuations beyond the mean field.  
Here we have first found that the effective mass of 
a composite fermion(CF) exhibits a curious step-function like 
behaviour against the Landau level 
filling.\cite{PRL84_03942_00,onodaPRB01}  
A most well-defined way to probe the interaction is to see whether 
Landau's Fermi liquid picture holds.  We have 
then examined this to conclude that the inter-CF interaction 
is of Hund's type (negative exchange) in both spin and 
orbital channels, and the strength of the 
interaction is significant (or can even be anomalous).

The strong-correlation limit raises an interesting 
question of how other quantum phases 
such as the BCS paired state should appear, especially 
as compared with ordinary correlated electron 
systems on lattice structures.  
For the latter, it is becoming increasingly clear 
that the shape of the Fermi surface controls the occurrence 
of pairing, which is anisotropic and includes the possibility 
of p, d, f symmetries\cite{MBT}.  
By contrast the FQH system is isotropic but the interaction 
(between the composite fermions) is 
controlled by the Landau index, for which 
we here examine the CF pseudopotential in the particle-hole 
channel from the viewpoint of the Fermi surface effect.\cite{oxford}  

Finally we ask ourselves: can such QHE physics in 2D 
systems have a possible extension 
to {\it three-dimensional} systems in strong magnetic fields? 
In 3D the kinetic energy, hence the shape of the Fermi surface, 
should become relevant.  
As a starting point I shall describe our recent proposal for 
the magnetic-field induced SDW in 3D,\cite{KoshFISDW} 
where the integer QHE, which 
is predicted by Koshino {\it et al} \cite{Koshino} to 
occur in 3D on the fractal energy spectrum 
similar to Hofstadter's, should be realised 
in entirely realistic magnetic fields.  The mechanism for the field-induced 
phase is an effect of interaction in Landau's quantisation 
on incompletely-nested Fermi surfaces (that become 
multiply-connected after the SDW formation), so 
the interference of the magnetic quantisation 
and the shape of the Fermi surface suggests an interesting future 
avenue for the many-body physics in 3D systems.  

\section{Effective mass of the composite fermion against $\nu$} 

There are several ways to define the mass of the composite fermion.   
The problem, 
related with the gauge-field fluctuations, is highly nonperturbative, 
where one way is to determine the mass and the interaction 
numerically from the (electron-hole) excitation spectra 
(i.e., a two-particle property) for finite 
systems.  So we have systematically studied even and 
odd fractions with spin degrees of freedom 
included.\cite{PRL84_03942_00,onodaPRB01}  
We have done this in two steps.  We first assume a 
free composite fermion picture to estimate the mass 
before moving on to Landau's Fermi liquid picture. 


For the FQH states at odd fractions, 
$\nu = 2\pi n_e/(eB) = d_s p/(d_s\tilde{\phi}p\pm 1)$, 
we estimate the effective mass, $m^*$, from an excitation gap, 
which is the effective cyclotron energy, 
$\omega_c^*=eB^*/m^*$ for $eB^* = eB - 2\pi \tilde{\phi}n_e$, 
in the free composite fermion picture with an 
effective $\nu^*\equiv 2\pi n_e/(eB^*)=d_s p$.  
Here $\tilde{\phi}$ is the number of flux quanta attached to 
each electron, $n_e$ the number density of electrons, $p$ a positive integer, 
$d_s=2$ the spin degeneracy ($d_s=1$ for the spinless case) 
and the natural units with $\hbar=c=1$ are adopted.  

For even fractions, $\nu=1/\tilde{\phi}$, 
we can estimate $m^*$ for a metal of composite fermions.  
In this case the low-lying excitation is 
(electron-hole pair) excitations around the 
``Fermi surface", and $m^*$ can be determined how the 
excitation gap vanishes for $N_e\rightarrow \infty$, 
when the number of electrons is 
$N_e=d_s (l_{\rm F}+1)^2$, i.e., the closed-shell case 
with $l_{\rm F}$ being the Fermi angular momentum.

Let us look at the effective mass thus 
estimated\cite{PRL84_03942_00,onodaPRB01} 
in the free CF picture for various values of Landau-level filling 
in Fig.2, which plots the 
inverse effective mass against $\nu$ for the spinless case 
for the $1/r$ Coulomb interaction.  
We can immediately see that the effective mass against $\nu$, while 
basically becoming heavier 
for $\nu=1/2 \rightarrow 1/4 \cdot \cdot \cdot$, 
exhibits a step-function-like behaviour, where each step corresponds to 
each number of attached flux quanta, $\tilde{\phi} (=2,4,..)$.  
Within each step, $m^*$ is only weakly dependent 
on $\nu$ (regardless of fractions odd or even), 
which implies that the effective mass 
is basically determined by $\tilde{\phi}$, the number of attached fluxes. 

This is totally unexpected, since the CF theory in a 
mean-field treatment predicts a smooth function 
(dashed line in the figure, which will discussed below)\cite{MSmass}.  

Within each step, , the number of attached fluxes. 

Note at the same time 
that the usual practise of linearly plotting $\hbar \omega_c^*$ 
versus $B^*$\cite{Duetal} is allowed only when 
$m^*$ is nearly constant in each region specified by $\tilde{\phi}$.  
Experimentally, the effective mass 
from the thermal activation energy in the Shubnikov-de Haas oscillation 
by Leadley {\it et al} shows a difference in the 
$B$-dependence between $\tilde{\phi}=2$ and $4$\cite{PRL72_01906_94}.  

\section{Composite fermion gas: a Fermi liquid?} 

Inclusion of the spin degrees of freedom turns out to vastly 
modify the excitation spectrum, which is a sign that 
the spin-spin (exchange) interaction between CF's is not negligible.  
The exchange interaction between composite fermions 
has been estimated for 
the spin wave, where the spin stiffness is shown to be 
larger as we change $\nu = 1/3 \rightarrow 1/5 \cdot \cdot 
\cdot$, which is explained by the composite-fermion picture 
(Fig.1; \cite{nakajima,nakajima95}).

Here we more closely look at the spin-dependent 
interaction in terms of the Fermi liquid picture for CF's 
for even-fraction metals.
This picture assumes that the excitation energy, $\delta E$, is given 
as a functional of the deviation in the particle number 
from the ground state 
with the Landau function 
$f_{\mbox{\boldmath{$l$}}\sigma\:\mbox{\boldmath{$l$}}'\sigma'}$ as a 
coefficient. 
We can then expand $f$ to the first order in
$\mbox{\boldmath{$l$}}\cdot\mbox{\boldmath{$l$}}'$ 
(which corresponds to the spherical harmonics expansion for 
$f_{\mbox{\boldmath{$p$}}\:\mbox{\boldmath{$p$}}'}$ 
in a flat system)  and
$\mbox{\boldmath{$\sigma$}}\cdot\mbox{\boldmath{$\sigma$}}'$.  
After a bit of algebra\cite{onodaPRB01}, we end up with 
\begin{eqnarray}
\delta E 
&=& 
\Delta_{\rm FL} 
\biggl[1+\frac1{d_s(l_F+1)}\Bigl\{(G_0+G_1)
\mbox{\boldmath{$S$}}\cdot\mbox{\boldmath{$S$}}
\nonumber\\
&&
+\frac14[F_1+G_1(3-
 2\mbox{\boldmath{$S$}}\cdot\mbox{\boldmath{$S$}})]
 \frac{\mbox{\boldmath{$L$}}\cdot\mbox{\boldmath{$L$}}}
{l_F(l_{\rm F}+1)}\Bigr\}\biggr],
\nonumber
\end{eqnarray}
where $\Delta_{\rm FL}=(l_{\rm F}+1)/(m^*_{\rm FL}R^2)$ with 
$m^*_{\rm FL}$ being the effective mass defined in the context of
the Fermi liquid theory, $F$'s and $G$'s are (dimensionless) Landau 
parameters, 
$\mbox{\boldmath{$L$}}=\mbox{\boldmath{$l$}}-\mbox{\boldmath{$l$}}'$
and $\mbox{\boldmath{$S$}}=(1/2)
(\mbox{\boldmath{$\sigma$}}-\mbox{\boldmath{$\sigma$}}')$ 
(the total angular momentum and spin, respectively, 
of the excitation from a closed shell).

So we can estimate $m^*_{\rm FL}$
and the Landau parameters {\it simultaneously} 
by best-fitting the numerical results.\cite{onodaPRB01}
First thing we notice is that the coefficient 
($G_0+G_1$) for the
$\mbox{\boldmath{$S$}}\cdot\mbox{\boldmath{$S$}}$ term
is {\it negative}, i.e., we have a Hund's second rule 
exchange interaction.  This is readily seen from Fig.3(b), 
where $\delta E$ for spin-flip excitations is 
significantly smaller (about one half, which implies that $G_0+G_1 \simeq -1$ 
for $N_e=8$). The observation is consistent with 
previous results.\cite{nakajima,matsue}

We can in fact examine the whole spectrum. 
If we take the spinless case for simplicity 
and look at the low-lying excitation spectrum 
(solid circles in Fig.3(a)) against the 
angular momentum $L$, which nuclear physicists would call 
an ``Yrast spectrum", 
the spectrum delineates the lower boundary 
of the Fermi liquid excitation spectrum.  The 
free CF result (crosses) is so good as to reproduce the spectrum, 
including the shell structure that has to do with $2nk_F$ effects, where 
$k_F$ is the Fermi wavenumber of the composite fermion.  

To be more precise, the exact result lies significantly below the 
free CF result for larger $L$.  Since in the Landau's picture 
the excitation energy contains a term 
$\propto F_1^{\rm spinless} 
{\mbox{\boldmath{$L$}}}\cdot{\mbox{\boldmath{$L$}}}$ 
with $F_1^{\rm spinless} = F_1 + 3G_1$, this implies 
the Landau parameter $F_1^{\rm spinless}$ is {\it negative}, 
i.e., CF's have an orbital exchange coupling 
of the Hund's first rule, as consistent previous works on 
the $\nu=1/{\rm even}$ ground state
\cite{PRL72_00900_94,Morf_d'Ambrumenil}.
So the FQH system has spin- and orbital-exchange couplings both of which 
are Hund's type.   
The Fermi-liquid result for $m^*_{\rm FL}$ is comparable with
$m^*$ estimated from the free CF picture.  

However, we have to hasten to add that the Landau parameter, 
$F_1^{\rm spinless}$, is ill-behaved in 
that the quantity is sample-size-dependent: 
$F_1=-0.5 (-1.0)$ for $N_e=9 (16)$.  
Landau parameters should be scale invariant in a Fermi 
liquid, so the anomalous behaviour should indicate 
that the ordinary Fermi liquid picture cannot be directly applied.  
This may have to do with the relation,
\[
m^*_{\rm FL}/m_{\rm bare} = 1+F_1^{\rm spinless}/2
\]
(with $1/2$ due to the dimensionality of two), where 
the infinite bare mass for the Landau level for 
$N_e \to \infty$\cite{divmass} would imply 
$F_1^{\rm spinless}\rightarrow -2$, an anomalously large value.

We have also investigated how the situation 
changes as the interaction is made shorter-ranged or longer-ranged 
than the Coulombic, since RPA\cite{RPA} and 
renormalisation-group results\cite{RG} suggest that 
the system is a marginal Fermi liquid just for the Coulombic 
interaction while a normal Fermi liquid recovered for shorter-ranged 
ones.  When the functional form of the interaction is 
made shorter-ranged [$V(r) \propto 1/r^{\alpha}; \alpha>1$], 
both $m^*_{\rm FL}$ and 
the Landau parameter $|F_1|$ increase 
($F_1 = -0.06 \rightarrow -0.8 \rightarrow -1.2$ 
for $\alpha=0.5 \rightarrow 1.0 \rightarrow 1.5$). 
In the thermodynamic limit, the 
Landau function could possibly be singular, 
as considered by Stern and Halperin\cite{PRB52_05890_95} 
by summing the diagrams in accordance with the Ward-Takahashi identity.  
The size-dependence of $F_1^{\rm spinless}$ may be related to 
the marginal Fermi liquid predicted with RPA in Ref.\cite{RPA}.
Note that we have deduced $m^*$ from 
the excitation spectrum of the lowest-Landau-level projected model,
i.e., a two-particle property in a model with an infinite bare mass.
So, even when the mass defined from the pole of the 
one-particle Green's function is anomalous, the 
excitation energy (e.g., the energy required to 
create a particle-hole pair) can be less anomalous.   
At any event, if the Fermi or marginal Fermi 
liquid persists in the thermodynamic limit, 
this will serve as an instance in which a system that 
has no small parameters 
(interaction/kinetic energy$=\infty$, $\tilde{\phi} \sim O(1)$) 
can be a Fermi liquid.

\subsection{Effect of gauge field --- Shankar-Murthy theory}
Let us further discuss the mass from the viewpoint of 
the gauge field fluctuations.  
The $1/m^* = (0.185 \pm 0.002) e^2\ell$ at $\nu=1/2$,
where ${\ell}\equiv 1/\sqrt{eB}$ is the magnetic length, 
obtained here from the excitation gap is
slightly smaller than $1/m^*\simeq  (0.2\pm0.02)e^2\ell$,
estimated from the ground-state energy per particle\cite{Morf_d'Ambrumenil}.
On the other hand the present value, at $\nu=1/2, 1/4, 1/6$, 
is curiously close to an 
analytic estimate, $1/m^*\simeq e^2\ell/6$, obtained from 
the version of the composite-fermion theory due to 
Shankar and Murthy that incorporates 
the effect of the correlation hole\cite{Shankar_Murthy,MSmass}.

There are various versions of the composite fermion theory.  
While the most naive one just attaches fluxes to an electron 
(a singular gauge transformation), this does not say anything 
about why the electron-electron repulsion requires this.  
Then Read introduced a physically clearer version, in which 
the electron correlation effect is nicely incorporated 
as the correlation hole attached to the CF transformation.  
The penalty for doing that is the transformation loses its 
unitarity.  Motivated by this, Shankar and 
Murthy\cite{Shankar_Murthy,MSmass} introduced 
a Hamiltonian theory, where the unitarity is recovered, but at the 
cost of a complexity, in which one first expands the Hilbert 
space, and then restrict it to a physical one.  The expanded space 
is called ``all-$q$" theory, the restricted one ``small-$q$".  
The result in the latter (dashed line in Fig.3) agrees with 
the present result, while the former has a vastly different result.

\subsection{Higher Landau levels and paired states}

There is a growing realization that the FQH system 
can be very sensitively affected when we go from the 
lowest Landau level ($N=0$) to higher ones
in the series $\nu=\nu^{(N)} + 2N$.  
Theoretically, the interaction between CF's 
strongly depends on the Landau level index.  
For $\nu=5/2 (\nu^{(1)}=1/2, N=1)$, which sits in between 
a Fermi-liquid and the stripe, trial functions for 
paired BCS states have been proposed.  
Specifically, a $p_x-ip_y$-wave, spin-triplet paring of CF's 
proposed for $\nu=5/2$ by Moore and Read\cite{NPB360_0362_91} 
is supported by numerical studies by Morf\cite{PRL80_01505_98}
and by Rezayi and Haldane\cite{PRL84_04685_00}, as well as by 
a recent experiment by Willet {\it et al}\cite{PRL88_066801_02}.  
However, paring mechanism has yet to be fully understood. 
We have seen that composite fermions are strongly 
interacting, and let us make two remarks here.

First question is how Hund's first and second rules, 
shown above for the lowest Landau level, will be modified 
in higher Landau levels.  Morf and d'Ambrumenil 
conjectured, from numerical results on the violation of the 
Hund's rule combined with the Rezayi-Read trial wavefunction, 
that the compressible state becomes unstable 
for $\nu^{(N)} = 1/\tilde{\phi} \geq 1/2N$.  
Our numerical result\cite{oxford} for finite FQH system 
for the total angular momentum, $L$, 
and the total spin, $S$ in the ground state 
also shows that the Hund first and second rules are 
obeyed ($\bigcirc$) as \par
\begin{tabular}{l|c|c|c}
 & $N=0$ & 1 & 2 \\
\hline
$\nu^{(N)}=1/2$ &$\bigcirc$ & &  \\
$\nu^{(N)}=1/4$ &$\bigcirc$ &$\bigcirc$ &  \\
$\nu^{(N)}=1/6$ &$\bigcirc$ &$\bigcirc$ & $\bigcirc$  \\
\end{tabular}
\ \\

Second point is a BCS trial function with $p_x-ip_y$ pairing 
due to Greiter, Wen, and Wilczek,\cite{PRL66_03205_91}
who studied the Moore-Read paired state
in the spherical geometry with a wavefunction,
\begin{eqnarray}
\Psi({u_i, v_i}) = {\rm Pf} \left[\frac1{u_i v_j - u_jv_i}\right]
\prod_{i<j}^{N_e}(u_i v_j- u_jv_i)^2, 
\nonumber
\end{eqnarray}
where $u_i$ and $v_j$ are the spinor variables for $i$-th electron and 
Pf stands for the Pfaffian.  A numerical result for 
the radial distribution function, $g(r)$, for $\nu=1/2$ in 
Fig.4 exhibits a significant difference between 
$N=0$ and $N=1$ Landau levels, where the latter is 
characterised by the fact that the inter-CF interaction in the 
particle-hole channel has a dip just around $k_F$ for the composite 
fermion\cite{oxford}. 
This kind of instability in a Fermi liquid reminds us of a theorem for the 
usual (zero $B$) electron gas due to Kohn and Luttinger\cite{PRL15_00524_65}, 
who showed that the normal liquid has to become, at low enough temperatures, 
unstable against anisotropic pairing, where the instability is associated with 
the Friedel oscillation in a system with a well-defined 
Fermi surface.

\section{Quantum Hall effect in three dimensions}

Having reviewed the quantum Hall physics for two-dimensional 
systems, one fundamental question is: 
while the QHE is usually conceived as specific to
two-dimensional(2D) systems, can QHE occur in three dimensions(3D), 
and, if so, how?  
It has been suggested that, {\it if} there is an energy gap with the Fermi
energy lying in it, integer QHE can occur in 3D 
accompanied by quantised Hall tensor components 
$\sigma_{ij}$\cite{Halp,Mont,Kohm}.  
Usual wisdom, however, is that gaps should disappear in 3D, 
since the motion along the magnetic field is basically free, 
and gaps should be smeared out.  
However, we have recently shown \cite{Koshino,Koshinoisotropic} 
that, under proper 
conditions, we do have gapful
energy spectra, which are fractal with recursive gaps as 
in the Hofstadter butterfly for 2D periodic systems in 
magnetic fields.  Interesting points are: 

(i) The integer QHE in 3D 
is by no means a remnant of the 2D butterfly, since the 3D butterfly 
is washed out when the third direction hopping is turned off.  

(ii) The butterfly appears in 3D when 
plotted versus the tilting angle of the magnetic field.  

(iii) The magnetic field required for the 
3D butterfly is as modest as $\sim 40$ T in anisotropic 3D system 
when we employ higher Landau levels for the butterfy\cite{Koshino}, 
which is orders of magnitude smaller 
than those required for the 2D butterfly.\cite{MPI}   

An appealing possibility, in the context of the 
present article, is: can we exploit the 
electron-electron interaction to realise the QHE in 3D?  
We in fact propose that the 
magnetic-field induced spin density wave (FISDW), considered in 3D for 
the first time, is promising.\cite{KoshFISDW}  While this is a mean-field 
theoretic effect unlike the FQHE that is a correlation effect, 
this would be a start.\cite{3DFQHE}  
The FISDW has originally been conceived for a 2D organic metal 
(TMTSF, called Bechgaard salt), where an anisotropic 2D Fermi surface is 
incompletely nested, 
and Landau's quantisation for the pocket arising from the 
SDW gap formation results in a series of gaps\cite{Gorkov,Poil,Maki}. 
The Bechgaard salt has the hopping 
integrals $t_x: t_y: t_z \sim 1: 0.1: 0.003$, where the third-direction
hopping $t_z$ happens to be so small that the
system is almost completely 2D.  
So our extension to 3D must answer the question: 
can a 3D-specific FISDW systematically result in a 3D IQHE?  
Some studies\cite{Mont_Litt,Hase95} 
have discussed a quantisation of $\sigma_{ij}$ 
when there is a 3D FISDW, but it remained to be clarified whether and how 
FISDW phases arise in 3D.
We show that, if we have an anisotropy $t_x \gg t_y
\approx t_z$ (as opposed to $t_x \gg t_y \gg t_z$ in TMTSF), 
a 3D butterfly spectrum, accompanied by the 3D QHE, should indeed 
occur. The phase diagram against the tilted magnetic field $(B_y,
B_z)$ is obtained by optimising the SDW nesting vector.  
The quasiparticle spectrum is fractal, and 
the QHE numbers (topological invariants) associated with 
the gaps are obtained.  

Remarkably, the butterfly and
quantised $(\sigma_{xy},\sigma_{zx})$ are intimately related with
the FISDW in that the quantum numbers characterising the nesting vector
coincides with the QHE numbers.  The QHE is again a {\it genuinely} 3D effect, 
since they go away if we turn off $t_z$.
We have also addressed the surface currents in 3D QHE, 
in analogy with the edge currents in 2D QHE.  
Interestingly, the 3D surface Hall conductivity 
{\it exactly} coincides with that given by the bulk Hall conductivity, 
just as the edge and bulk Hall conductivities coincide in the 2D QHE. 

\subsection{Magnetic field induced SDW in 3D}

We consider a simple orthorhombic metal with an energy dispersion
$\epsilon(\Vec{k}) = -t_x \cos k_x a -t_y \cos k_y b -t_z \cos k_z c,$
where $a,b,c$ are lattice constants and the 
transfer energies are assumed to satisfy $t_x \gg t_y, t_z$.
The dispersion along the conductive $k_x$ can be 
approximated around the Fermi energy as a linear function, 
$v_F (|k_x| -k_F)$, 
and the three-dimensionality (warping of the Fermi surface 
due to $t_y$ and $t_z$) is considered. 
We apply a magnetic field $(0,B_y,B_z)$ normal to the conductive axis $x$, 
and examine the SDW order parameter $\Delta(x)$ in a 
mean-field equation for the wavefunction 
with the 3D nesting vector $\Vec{q}= (q_x,q_y,q_z)$, which can be written as
\begin{eqnarray}
&&
\left(
\begin{array}{cc}
E - H_{\uparrow}(x) & \Delta(x) \\
\Delta^*(x) & E - H_{\downarrow}(x)
\end{array}
\right)
\left(
\begin{array}{c}
u(x) \\ v(x)
\end{array}
\right)
 = 0 ,
\nonumber \\
&&H_{\uparrow}(x) = 
-i v_F\partial_x + \epsilon_\perp(\Vec{k}_\perp + e\Vec{A}_\perp(x)),
\nonumber \\
&&H_{\downarrow}(x) = 
+i v_F\partial_x + \epsilon_\perp(\Vec{k}_\perp - \Vec{q}_\perp + e\Vec{A}_\perp(x)),
\nonumber
\label{schrodinger}
\end{eqnarray}
where $\Vec{k}_\perp \equiv (k_y,k_z)$, 
$\Vec{A}_\perp(x) = (B_z x, -B_y x)$ is the vector potential, 
$H_{\uparrow} (H_{\downarrow})$ the Hamiltonian for
an electron on the right Fermi surface with up-spin 
(or on the left Fermi surface with down-spin) with 
$u (v)$ being corresponding wavefunctions.
The SDW gap $\Delta(x)$ can be approximated by a single-mode, 
$\Delta(x) \sim \Delta e^{iq_x x}$, and we can 
determine $\Delta$ and $\Vec{q}$ self-consistently so as to minimize 
the ground state energy. 
The Fermi energy lies in the largest gap to minimize the energy, 
from which we can determine 
the SDW nesting vector with $q_x = 2k_F - M G_b - N
G_c$ with $G_b = eB_zb, G_c = eB_yc$.

The situation here is reminiscent of (or indeed mathematically 
the same as) in the 3D
butterfly studied more generally in Ref.\cite{Koshino}.  
The physics is the following.  In the ordinary Hofstadter's 
butterfly in 2D, interference of 
Landau's quantisation and Bragg's reflection (band gap) 
gives the butterfly.  In the butterfly in 3D,\cite{Koshino} 
two Landau quantisations 
(on $x$-$y$ and $z$-$x$ planes) in tilted magnetic fields $(B_y,B_z)$
interfere, which gives rise to the fractal 
energy spectrum.  
In the FISDW in 3D, the two components of the nesting wavenumber, 
$G_b(\propto B_z), G_c(\propto B_y)$, interfere.  
This reasoning dictates that the spectrum
plotted against $B_z/B_y$ should have a structure similar to 
Hofstadter's butterfly, which is indeed the case as shown in Fig.5.  
Another way of explanation 
is that the Fermi surface in both cases (i.e., anisotropic 
3D crystals and incompletely nested 
Fermi surfaces) have {\it multiply-connected} structure, on which 
Landau's quantisation takes place.  

Figure 6(a) shows the phase diagram, 
plotted against $B_y$ and $B_z$, 
where two integers represent the QHE integers.  
The quasi-particle spectrum plotted against $B_z/B_y$ in
Fig.6(b) has indeed a fractal structure. 
An important difference, however, from the butterfly in the 
non-interacting case is that the SDW phase, being interaction-originated, 
{\it adjust 
itself} in such a way that the largest gap in the butterfly has the Fermi
level in it.  The set
of integers $(M,N)$ that give the largest gap vary in a complicated
sequence as the field is tilted, where the $(M,N)$ have an important
physical meaning --- the Hall conductivity.  Following Yakovenko's
formulation for 2D, Sun and Maki\cite{Sun} have predicted that the Hall
conductivities in the FISDW with $(M,N)$ are given by 
$(\sigma_{xy}, \sigma_{zx}) = -(e^2/h)(M/c,N/b)$ (per spin). 
In \cite{Koshino} we have obtained the QHE integers for the 3D butterfly
using the general Widom-St\v{r}eda argument\cite{widomstreda} due to
Halperin, Kohmoto and Wu\cite{Kohm}, where these integers are identified as 
topological invariants assigned to each gap in the butterfly.

\subsection{Experimental feasibility}

The magnetic field required for the 3D FISDW is dramatically 
reduced to $B \sim 10$ T (for the magnitude of transfer integrals 
typically found in organic metals), so this should be entirely 
within experimental feasibility.\cite{sheathMo} 
Another novel candidate should be 
doped {\it zeolites}.  It has been established that guest 
atoms such as potassium can be incorporated into the 
nanometer-sized cages in zeolites,\cite{nozue} 
where the electronic structure is shown to be 
surprisingly simple.\cite{zeolitePRL}  So an application of magnetic field to 
zeolites (preferably anisotropic ones such as ZSM-5)
will be interesting. 
We can also apply two external modulations (such as the acoustic waves) 
to the otherwise uniform system to realise the long periodicity.\cite{EP2DS}

\subsection{Wrapping current in the QHE in 3D}

Having looked at the bulk property,
let us consider the surface states in 3D QHE, 
since the edge states\cite{Halp1982} are an important issue in the 2D QHE.  
Koshino, Halperin and the present author\cite{Kosh_wrap} have shown that 
the 3D QHE in a finite sample should accompany a {\it wrapping current} 
that winds around the faces of a 3D sample.   

Curiously, the current direction on each facet does not coincide with 
the plane normal to the magnetic field $\Vec{B}$, but 
is dictated by the 3D topological (Chern) integers, 
which are just the quantised Hall tensor components, 
($\sigma_{yz},\sigma_{zx},\sigma_{xy})$ in 3D, 
since the Hall current 
$\Vec{j} = -(\sigma_{yz},\sigma_{zx},\sigma_{xy})\times\Vec{E}$ 
in an external electric field $\Vec{E}$. 
So this is a hallmark of the 3D-specific nature.  We can also show 
that the 3D Hall conductivity when all the currents are 
assumed to be carried by the wrapping current 
{\it exactly} coincides with that given by the bulk Hall conductivity. 
This is shown again by Widom-St\v{r}eda formula\cite{widomstreda} combined 
with thermodynamic Maxwell's relation.  
In 2D the Hall current carried by the edge current 
coincides exactly with one calculated with the Kubo formula\cite{AokiAndo} 
for the 2D sample, as has been shown by Hatsugai\cite{Hats} 
by identifying the connection between the topological (Chern) integers
for the bulk and the edge states.
So this property remarkably extends to 3D. 
We can in fact give an intuitive 
way to understand why surface or bulk does not really matter.\cite{Kosh_wrap}   

We can also propose an experiment to detect the 3D integer quantum Hall effect 
through the wrapping current.  
To observe the currents we have to attach some electrodes, 
and how to measure the conductivity tensor experimentally in the 3D QHE 
becomes much less trivial than in 2D.  
In analogy with the 2D Hall bar experiment 
we can attach two pairs of electrodes as shown in Fig.7, 
for which $\sigma^{\rm surface} =\sigma^{\rm bulk}$ can again be demonstrated.  

One interesting observation is the following.   
In the 2D Hall bar geometry it has long been recognized that there are
two ``hot spots" where the chemical potential has to drastically drop 
dissipatively.  In our 3D geometry, the hot spots
extend into two ``hot lines'' as shown in Fig.7.  
In 2D cyclotron emission has been observed around the hot 
spots\cite{kawano}, 
so we may be able to extend this to the hot lines.

As a final comment, formation of the plateaus, which has been 
explained in terms of the localisation due to disorder 
by Aoki and Ando,\cite{AokiAndo} 
is an interesting problem for the FQHE in 2D and 
QHE in 3D.  For the FQHE this is a challenge for the 
composite-fermion picture.  For the field-induced SDW, 
the system adjust itself in such a way that $E_F$ always sticks to the gap, 
which should act to widen the plateaus. 

\section{Final remarks}
To summarise, we have discussed 
(i) how the 2D continuous system in magnetic 
fields has interesting quantum phases arising from the interaction 
that is controlled by the Landau index, while (ii) in 3D 
systems the effect of the shape of the Fermi 
surface\cite{Kramerfest} 
enters as a novel ingredient in the physics in magnetic fields.  
If we combine (i) and (ii), even richer physics may be expected.  
Incidentally, 
in the context of the superconductivity and magnetism in 
heavy fermion compounds, a magnetic field induced triplet 
superconductivity has been proposed.\cite{ferrosuper}  Also, 
it has long 
been known and intensively studied that there is a 
rich phase diagram for liquid $^3$He that includes 
nonunitary pairing superfulid phases, so the combination 
of (i) and (ii) as conceived here may be related with, 
or possibly even go beyond, these.  
There are thus a wealth of 
open questions to be unraveled in the physics of quantum Hall 
effect even after the two decades of its discovery.  

The works described here are collaborations with Tatsuya Nakajima, 
Masaru Onoda, Takahiro Mizusaki, Takaharu Otsuka, Mikito Koshino, 
Toshihito Osada, Seiichi Kagoshima, Kazuhiko Kuroki and Bertland I. Halperin.

\newpage
Fig. 1 (a) The excitation spectrum for a finite FQH system, where 
the low-lying mode is the spin wave ({\large $\circ$}: exact result, 
$+$: composite-fermion result) for $\nu= 1/3, 1/5$.  
(b) How the Coulomb pseudopotential against the relative 
angular momentum $m$ changes as we attach the fluxes to electrons 
to convert them into CF. (after \cite{nakajima}).   

Fig. 2 Inverse effective mass estimated 
        from the size scaling of the excitation energy 
        in the spinless system ({\large $\bullet$}). 
        The dashed (dotted) line is the small-$q$ (all-$q$) 
        result in the Shankar-Murthy 
        CF mean-field. (after \cite{PRL84_03942_00,onodaPRB01}).

Fig. 3 
(a) Low-lying excitation spectra for $N_e = 16$ 
        in the exact diagonalisation ({\large $\bullet$}),
free composite fermion gas model ($+$) 
and the Fermi liquid theory (square) at $\nu=1/2$.  
(after \cite{onodaPRB01}).  Arrows indicate the Hund's coupling.
(b) Excitation gap ($\propto$ inverse effective mass) estimated 
        from the size scaling of the energy 
        for spinflip ({\scriptsize $\triangle$}) or 
        no-spinflip ({\large $\circ$}) excitations. 
        (after \cite{PRL84_03942_00}).

Fig.4 Radial distribution function $g(r)$ 
against the great-circle distance $r$ 
for $N_e^{(1)}=9$ system with $N_\phi^{(1)} = 16$.
Results for $N=1, N=0$ Landau levels, and that for a 
deformed pseudopotential to remove the dip at $k_F$ are shown.
(after \cite{oxford}).

Fig.5 (a) the energy spectrum of an anisotropic 3D system 
against the angle $\theta$ of the applied magnetic field.  
Pairs of numbers for each gap represent the 
Hall conductivity $(\sigma_{xy},\sigma_{zx})$ in units of 
$-e^2/ah$.  (b) The Hofstadter butterfly in 2D.  
Bottom insets depict sample geometries.

Fig.6 (a) Phase diagram for the FISDW in 3D 
is shown against $(B_y,B_z)$.  The
phases are labelled by the quantum Hall integers $(\sigma_{xy},\sigma_{zx})$ in units of $(h/e^2)$.  The 3D-natured phases (which 
vanish for $t_z\rightarrow 0$) are shaded.
(b) The quasi-particle energy spectrum
against $B_z$ with a fixed $B_y$.
Vertical lines indicate boundaries between 
different FISDW phases labelled by $(\sigma_{xy},\sigma_{zx})$.
(c) An incompletely nested Fermi 
surface, which resembles, after the SDW gap formation, the 
multiply-connected Fermi surface for anisotropic 3D systems.

Fig.7 Right: Wrapping current (thin arrows) in the 
QHE in 3D (b), where the experimental setup of the electrodes 
to detect 3D QHE wrapping current is indicated.  
The corresponding picture in 2D is shown in (a).  
The ``hot lines (spots)'' are indicated by blurred lines (spots).  



\end{multicols}

\begin{thebibliography}{9}
\bibitem{divmass} 
This is not to be confused with the logarithmically diverging mass 
in Ref.\protect\cite{RPA}, 
where the effective mass is defined for 
the one-particle Green's function, not for 
excitations (a two-particle property) as is done here.
\end{thebibliography}
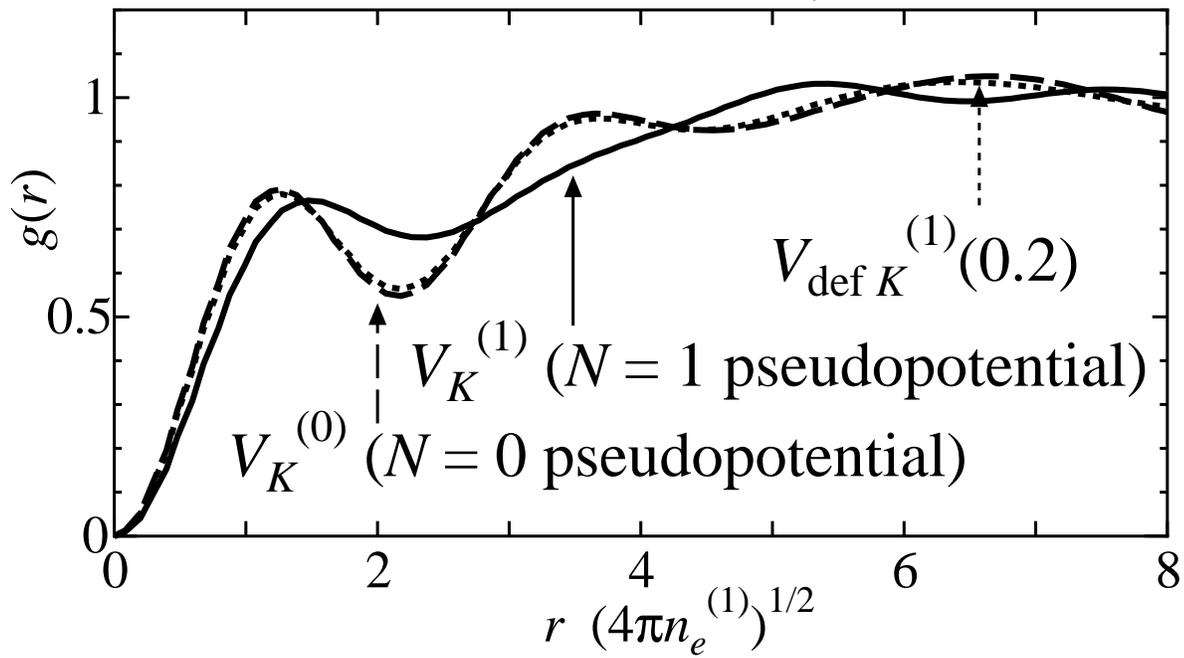
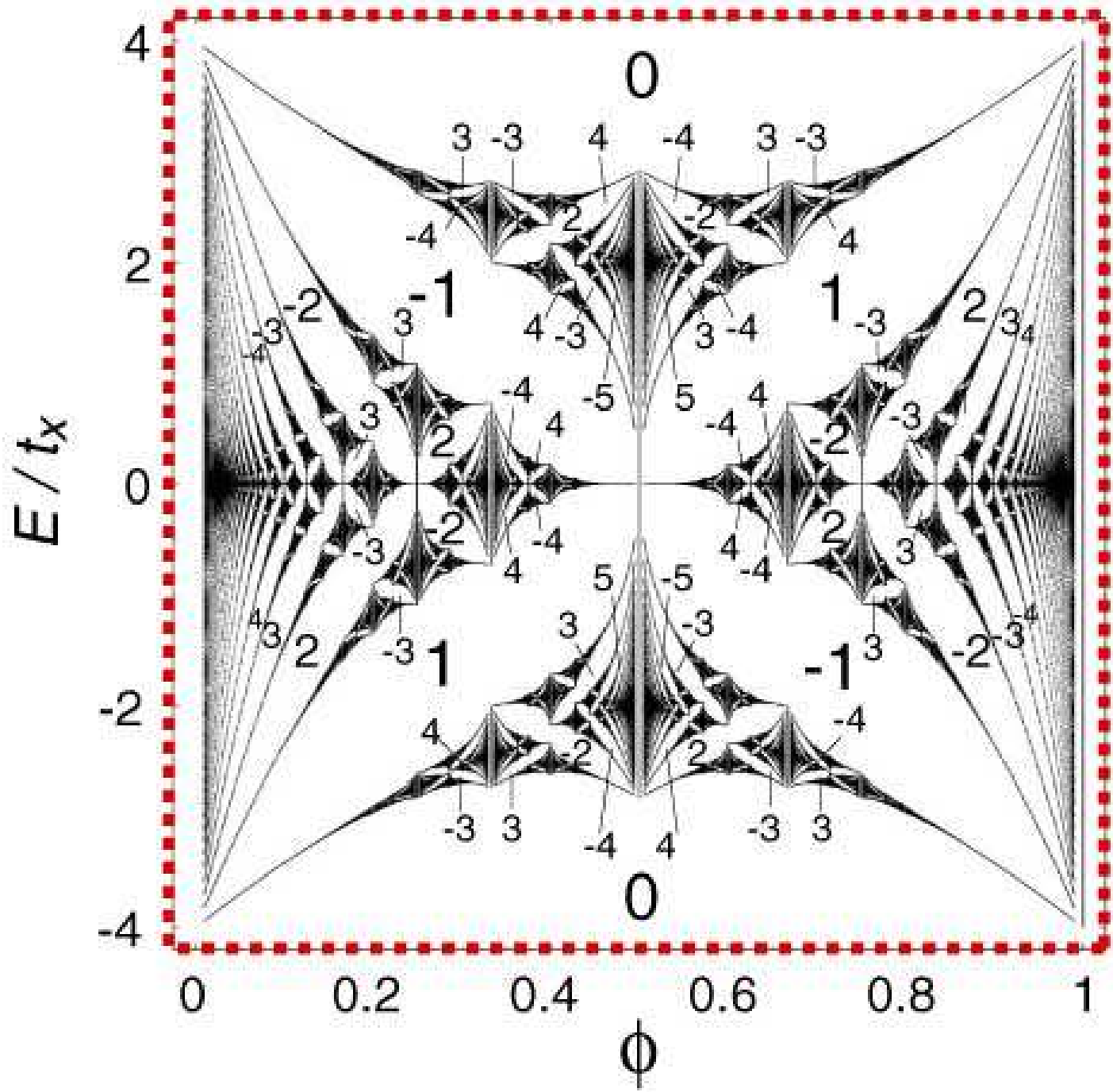
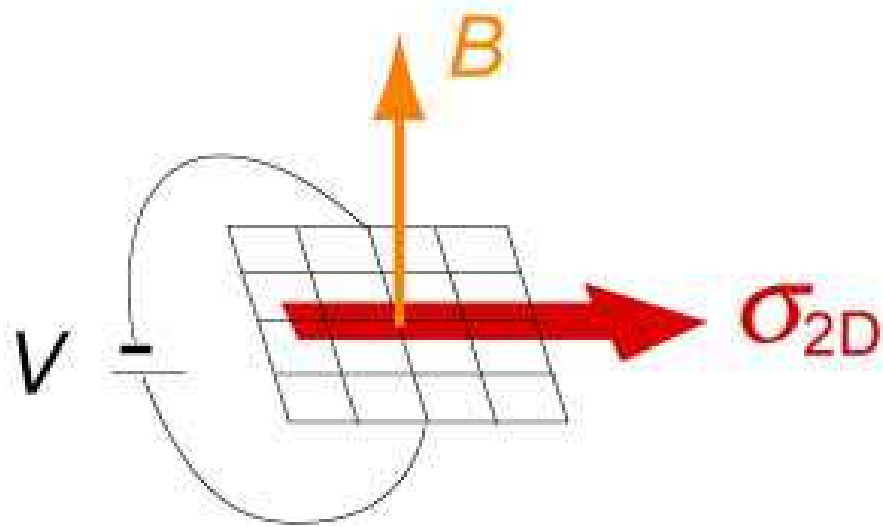
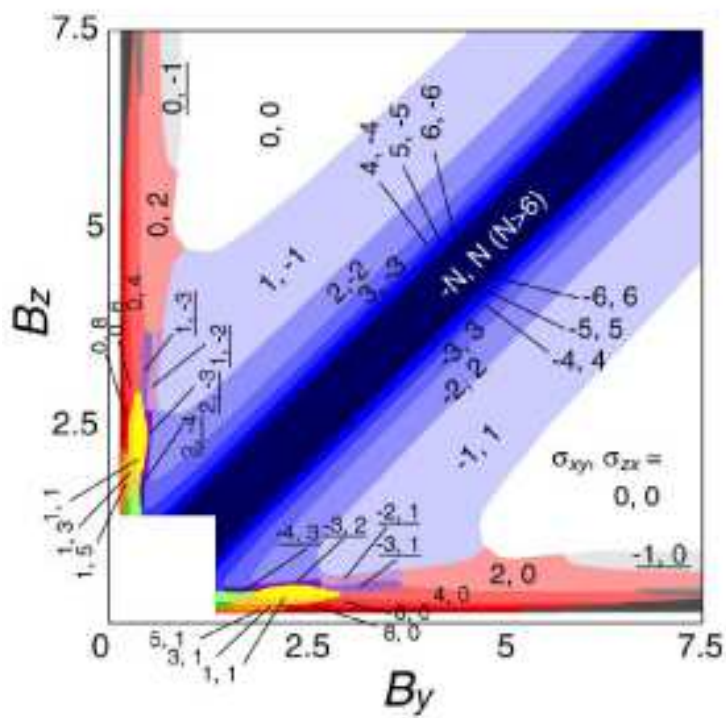
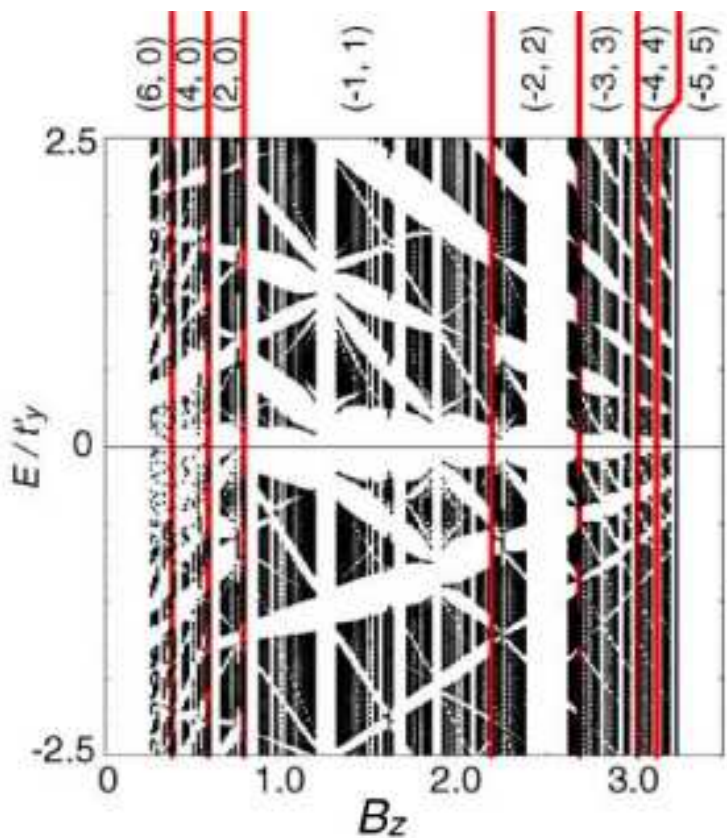
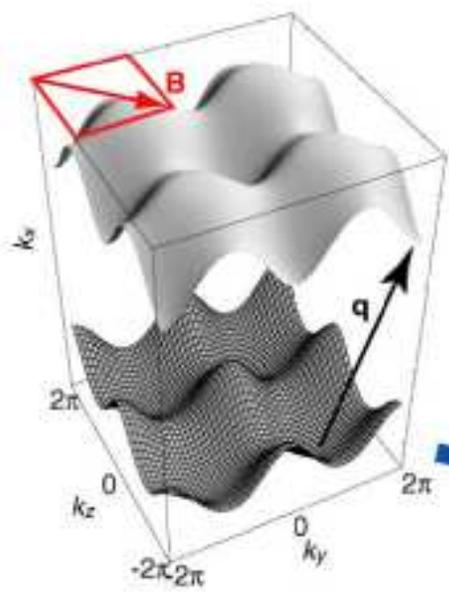
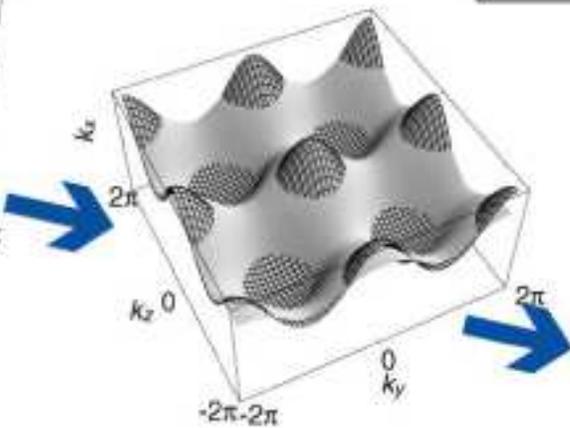
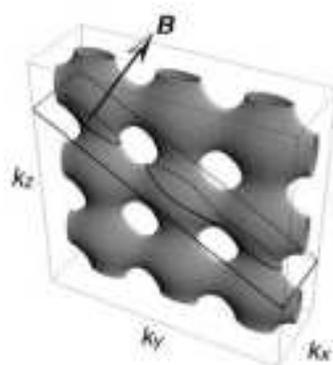
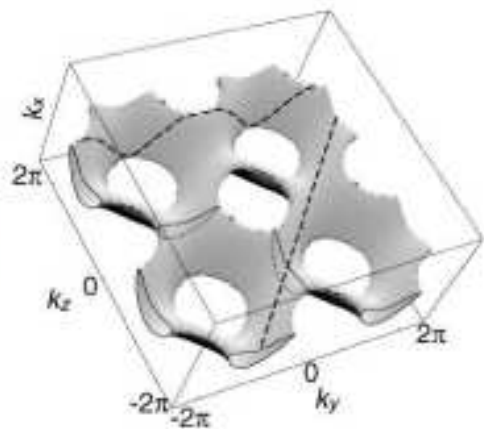
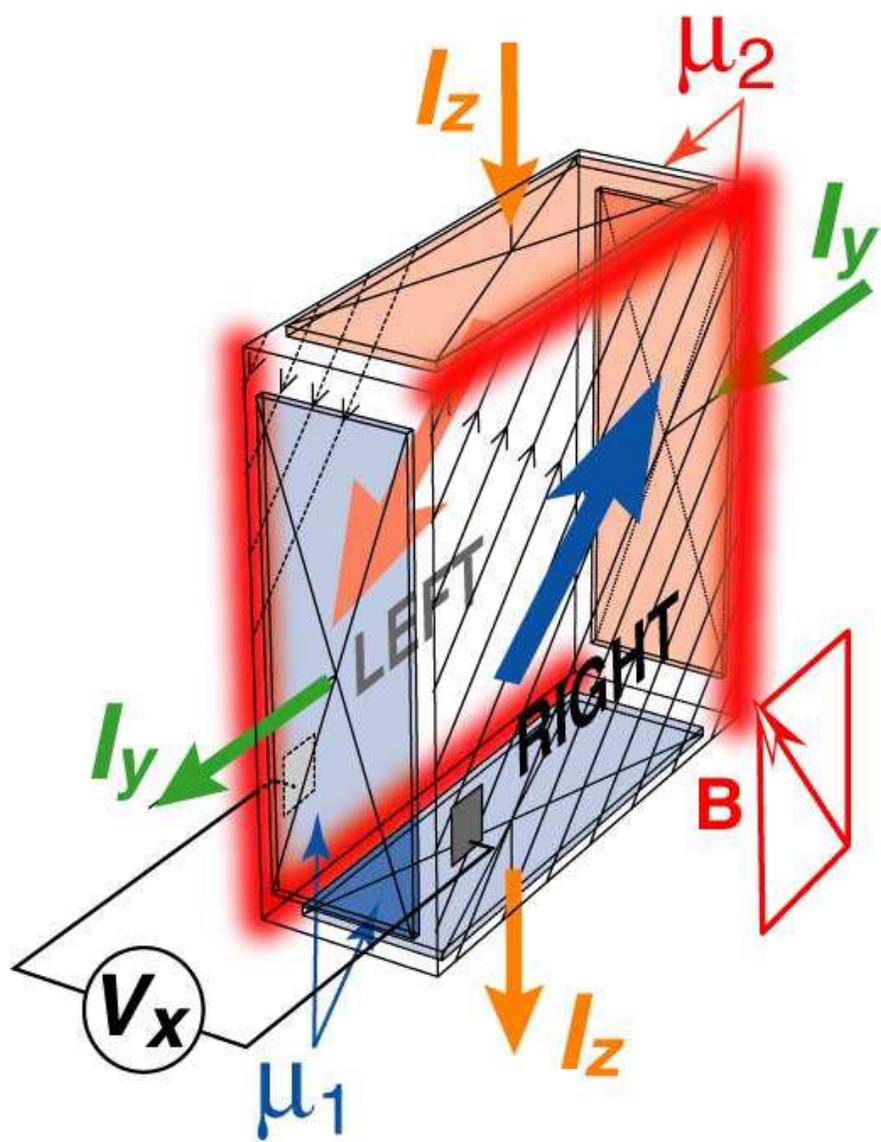
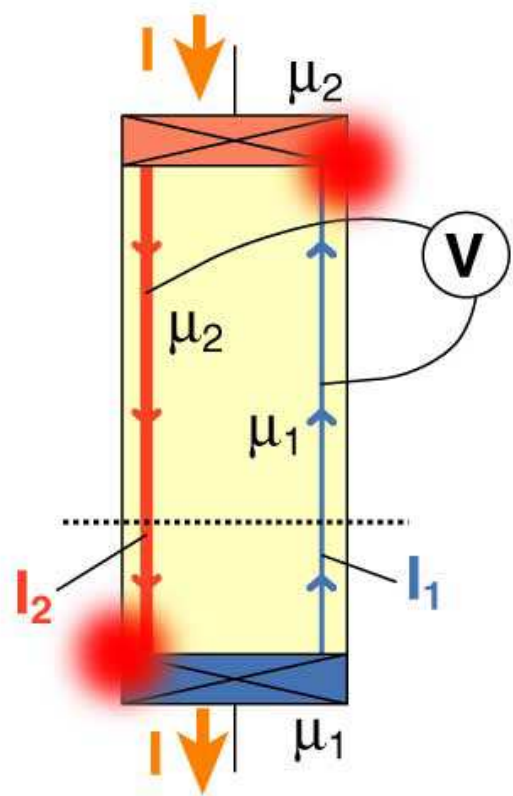
\end{document}